\documentclass{article}
\usepackage{graphicx}
\usepackage{amsmath}
\usepackage{amsfonts}
\usepackage{amssymb}
\usepackage{bm,bbm}
\begin{document}
\title{On the structure of the body of states
\\ with positive partial transpose}

\author{Stanis{\l}aw J. Szarek$^{1,2}$, Ingemar Bengtsson$^{3}$,
and  Karol \.Zyczkowski$^{4-6}$
\smallskip \\
$^1${\small Case Western Reserve University, Cleveland, Ohio, USA} \\
$^2${\small Universit{\'e} Paris VI, Paris, France} \\
$^3${\small Fysikum, Stockholm University, Stockholm, Sweden}\\
$^4${\small Perimeter Institute, Waterloo, Ontario, Canada}\\
$^5${\small Institute of Physics, Jagiellonian University, 
Krak{\'o}w, Poland}\\
$^6${\small Center for Theoretical Physics, Polish Academy of Sciences,
Warsaw, Poland}\\
}


\date{December 2, 2005}

\maketitle

\begin{abstract}
We show that the convex set of separable mixed states of the
$2 \times 2$ system is a body of a constant height.
This fact is used to prove that the probability to find a
random state to be separable equals 2 times the probability
to find a random boundary state to be separable,
provided the random states are generated uniformly with respect to
the Hilbert--Schmidt (Euclidean) measure.
An analogous property holds for the set of positive-partial-transpose
states for an arbitrary bipartite system.
\end{abstract}


The phenomena of quantum entanglement
became crucial in recent development of
quantum information processing.
In general it is not easy to decide
whether a given mixed quantum state is
entangled or separable \cite{Gu04}.
The situation gets simpler for the qubit--qubit and the qubit--qutrit
systems, for which it is known that a state
is separable if and only if its partial transpose
is positive \cite{Pe96,HHH96a}. However, even in the
simplest case of the two--qubit system, the structure of the
set of separable states is not fully understood.
This $15$-dimensional convex set ${\cal M}^{(4)}_S$
is known to contain the maximal ball inscribed
in the set of all mixed states ${\cal M}^{(4)}$
\cite{ZHSL98}.
Although some work has been done to estimate the volume of the set of separable
states \cite{GB02, Sl99,Zy99,Sza04,AS05} and to describe its geometry
\cite{KZ01,GMK05}, the exact volume of the set of separable states
is unknown even in this simplest case \cite{Sl04}.

In order to elucidate properties of
the set of separable states Slater studied the probability that
a random state be separable, inside the set of mixed states and at its
boundary, using the Hilbert--Schmidt measure. For the two--qubit
system he found numerically that the ratio $\Omega $ between these
two probabilities is
close to 2 \cite{Sl05a,Sl05b}.
The fact that it is more likely to find a
separable state {\sl inside} the total set of mixed states
than on its boundary is consistent with the
existence of the separable ball centred at the maximally
mixed state.

We prove that for this system
the ratio $\Omega$ {\sl is} indeed equal to 2, provided
we work in the Euclidean geometry and use
the measure induced by the Hilbert--Schmidt distance.
Furthermore, we show that the space of two--qubit (or qubit-qutrit)
separable states can be decomposed into pyramids
of the same height. These results hold also
for the convex set of states with positive partial transpose
(PPT states) for any $K \times M$ bipartite system.

We will need some generalities about convex sets.
Let  $X \in {\mathbb R}^D$ be a convex body in ${\mathbb R}^D$ (in particular
it is a closed, bounded set with nonempty interior)  and $\partial X$ its
boundary. Let $V$ denote the volume of
$X$  and $A$ the ($D-1$-dimensional) area of $\partial X$,
both computed with respect to the standard Euclidean geometry.
For each body we define the dimensionless ratio
\begin{equation}
{\gamma}(X) :=
\frac{ A(X)\,r(X)}{ V(X)}  \ ,
\label{gammadef}
\end{equation}

\noindent where $r=r(X)$ is the radius of the maximal ball inscribed
inside  $X$. This ball is known as the insphere. Before we see
why ${\gamma}(X)$ is interesting, we invoke some geometric definitions.
A {\sl face} of a convex body $X$ is the intersection of $X$ with an
affine hyperplane which intersects $\partial X$, but does not
separate (the interior of) $X$ into two pieces. Such  an affine plane is
known as a {\sl supporting hyperplane}.  Such hyperplanes are
important because any convex body equals the
intersection of the half-spaces defined by its supporting hyperplanes.
The  {\sl polar} $Y^\circ$ of a set $Y \subset {\mathbb R}^D$
is defined as  the set \
$\{z : \langle z, y \rangle \leq 1$ for all $y \in Y \}$. For additional
background information consult the literature \cite{Rockafellar, Gruber}.

We can now formulate a lemma which is stated for bodies whose insphere
is the unit ball (i.e., the ball of unit radius centred at
the origin). However, this hypothesis is just a matter of
convenience and can always be achieved by proper re-centring and
re-scaling. Note that while, in general, there may be
more than one inscribed ball with maximal radius, this ambiguity
cannot occur for bodies verifying the conditions of the Lemma.

\smallskip

\noindent
{\bf Lemma 1.}  {\sl For a convex body $X$ whose insphere is the unit ball
the following three conditions are equivalent:}

\smallskip

\noindent
{\bf 1.} {\sl Every  boundary point of $X$ is contained in a face tangent
to the inscribed ball.}

\smallskip

\noindent
{\bf 2.} {\sl $X = Y^\circ$, where $Y$ is a closed subset of the unit
sphere such that the convex hull of $Y$ contains the origin in its
interior.}

\smallskip

\noindent
{\bf 3.} {\sl ${\gamma}(X)= A/V=D$.}
\smallskip

\noindent Before we prove this, let us observe that the equality $A/V=D$,
from condition {\bf 3}, is  Archimedes' formula for the volume of a
$D$-dimensional {\it pyramid} of unit height---where  we define a pyramid
as the convex hull of the base and the apex of some  convex cone.  Hence
it will be true that
${\gamma}(X) = D$ whenever the body  can be decomposed into a union of
disjoint pyramids of fixed height with apex at the center of the
inscribed ball. This is  true for regular simplices, cubes and, more
generally, for any polytope circumscribed around the unit ball, and hence
verifying condition {\bf 1}. The equality $A/V=D$ holds also for balls,
which can be explained by the ball being a limit of an infinite sequence
of polytopes. Thus condition {\bf 1} can always be thought of as asserting
such ``decomposability into pyramids", and we will refer to a body obeying
it as a {\sl body of constant height}.
However, whenever the bases of pyramids (faces of $X$) are of dimension
lower than $D-1$, the limiting procedures needed to prove equivalences
will  require some care. (For the body of $N \times N$ density matrices
$D = N^2 -1$, while the maximal faces have dimension $D - 2N + 1$.)

\noindent {\it Proof of Lemma 1:} ${\bf 1} \Rightarrow {\bf 2}$:  We first note
that
$X$ equals the intersection of the half-spaces limited by supporting
hyperplanes corresponding to faces tangent  to the unit ball. Consider one
such supporting hyperplane.  Since it contains a tangent face, it
has a common point  with the unit sphere. On the other hand, it cannot cut the
unit ball in two, since it is  supporting to $X$ and the unit ball is
contained in $X$. Hence the  supporting hyperplane itself is tangent to the
unit ball, and so the corresponding half-space is of the form
$\{z : \langle z, y \rangle \leq 1\}$ for some  $y \in S^{D-1}$.
The set $Y$ is the closure of the set of points $y$ that arise this way.
The condition on the convex hull of $Y$ will be automatically satisfied as it
is equivalent to boundedness of $X$.

${\bf 2} \Rightarrow {\bf 1}$: First, if $Y$ is a subset of the unit sphere,
then the unit ball is inscribed into the polar of $Y$ with every $y \in Y$
a contact point. A point lies on the boundary of $X=Y^\circ$  if and only if
one of the inequalities that define $X$ as a polar body becomes
an equality, i.e., if it belongs to the face
$F_y := \{ x \in X : \langle x, y \rangle = 1 \}$ for some $y \in Y$.
$F_y$ is clearly contained in a hyperplane tangent to the unit ball and it
contains the point of tangency (namely, $y$), so it is itself tangent to the
unit ball. We have thus shown that every point of $\partial X$ belongs to a
face tangent to the unit ball, as required.

${\bf 1} \Rightarrow {\bf 3}$. With an insphere ${\bf S}^{D-1}$ of radius
1 we can regard any one of its points $\omega \in {\bf S}^{D-1}$ as
a unit vector, and we can define a real valued function $r(\omega)$ such that
$r(\omega) \omega \in \partial X$. Then

\begin{equation} V(X) = \int_{{\bf S}^{D-1}}d\omega \int_0^{r(\omega)}dr
\ r^{D-1} = \frac{1}{D}\int_{{\bf S}^{D-1}}d\omega \ r(\omega)^D
\ , \end{equation}

\noindent where $d\omega $ is the usual measure on the unit sphere. Similarly,
if $n(\omega)$ is the unit normal of $X$ at the point lying ``above'' $\omega$,

\begin{equation} A(X) = \int_{{\bf S}^{D-1}}d\omega \ \frac{r(\omega)^{D-1}}
{\langle \omega, n(\omega)\rangle } \ . \end{equation}

\noindent Here, and below, we rely on the fact that the normal
of a convex body is uniquely
defined, and continuous, almost everywhere.
Whenever the normal $n(\omega)$ is uniquely determined, the face containing
$r(\omega) \omega$ is also uniquely determined and hence perpendicular to
$n(\omega)$. By assumption ${\bf 1}$, that face, and the supporting hyperplane
containing it, are tangent to the unit ball. This leads to the equality
\begin{equation} \langle r(\omega) \omega, n(\omega)\rangle = 1
\label{3} \end{equation}

\noindent or $\langle  \omega, n(\omega)\rangle = 1/r(\omega)$; substituting
this into (3) and comparing with (2) we obtain
$V(X) = A(X)/D$.

${\bf 3} \Rightarrow {\bf 1}$. We first note that (by an elementary argument
using convexity) the set $\Xi := \{ \omega \in {\bf S}^{D-1} : 
r(\omega) \omega$
lies on a face of
$X$ tangent to the unit ball$\}$ is closed. If assumption ${\bf 1}$ does
not  hold,  the complement of $\Xi$ in  ${\bf S}^{D-1}$ is nonempty and
it is an open subset of this sphere, hence of positive measure. By
the same argument as in the preceding proof we have, almost everywhere
in ${\bf S}^{D-1} \setminus \Xi$,

\begin{equation} \langle \omega, n(\omega)\rangle > 1/r(\omega) \end{equation}

\noindent while still  $\langle \omega, n(\omega)\rangle = 1/r(\omega)$ for
$\omega \in \Xi$. Hence, from (2) and (3),

\begin{equation} A(X) < \int_{{\bf S}^{D-1}}d\omega \ r(\omega)^D =
D V(X) \ . \end{equation}

\noindent This concludes the proof. \hfill $\square$

\smallskip We can now easily establish

\smallskip \noindent
{\bf Lemma 2.}  {\sl  Any intersection of two bodies of constant
height $X_1, X_2$,
containing the same inscribed sphere, is a body of constant height}.
\smallskip

\noindent {\it Proof:} Let $Y_j$ be the sets defined by $X_j=Y_j^\circ$ for $j=1,2$.
Then $X_1 \cap X_2 = (Y_1 \cup Y_2)^\circ$
and of course $Y_1 \cup Y_2$ is a subset of the sphere, if each of the
$Y_j$'s was. Making use of property 2 we conclude that
$X_1 \cap X_2$ is a body of constant height. \hfill $\square$
\medskip

We are going to use the above geometric
concepts to investigate the set ${\cal M}^{(N)}$ of density matrices acting in
   $N$-dimensional Hilbert space. An operator
$\rho:{\cal H}_N\to{\cal H}_N$
   belongs to  ${\cal M}^{(N)}$ if it is Hermitian, $\rho=\rho^{\dagger}$,
is (semi) positive definite, $\rho \ge 0$, and is normalized, Tr$\rho=1$.
Due to the latter normalization condition the  set  ${\cal M}^{(N)}$
is compact and has dimensionality $D=N^2-1$.
The  sphere inscribed in  ${\cal M}^{(N)}$ is centred at the maximally mixed
state $\rho_* = {\bf 1}/N$ and has radius  $r=1 / \sqrt{(N-1)N}$, if
computed with respect to the Hilbert--Schmidt
distance, $d^2_{\rm HS}(\rho,\sigma)={\rm Tr}(\rho-\sigma)^2$.
Observe that $r$ is equal to the radius of the sphere inscribed inside
the $(N-1)$-dimensional simplex ${\Delta}_{N-1}$ with edge lengths
$\sqrt{2}$.

A recent analysis  \cite{ZS03} of the volume and area of the set of mixed
states shows that their ratio $A/V$ is equal to $\sqrt{N(N-1)} \, (N^2-1)$.
Therefore the coefficient  $\gamma$ reads then
\begin{equation}
\gamma({\cal M}^{(N)}) =
     r\,\frac{A}{V} = \frac{1}{\sqrt{(N-1)N}} \,
     \sqrt{N(N-1)} \, (N^2-1) = N^2-1 =D \: .
\label{muMN}
     \end{equation}

\noindent
This observation inspired us to propose

\smallskip

\noindent
{\bf Proposition 1.} {\sl The set ${\cal M}^{(N)}$ of mixed quantum states is
a body of constant height.}
\smallskip

\medskip \noindent Indeed, it is immediate that condition ${\bf 1}$ of Lemma 1
holds. A density matrix lies on the boundary of ${\cal M}^{(N)}$ if and
only if it has
a zero eigenvalue. Such a matrix belongs to the face consisting of all density
matrices with support in the subspace of the Hilbert space that is orthogonal
to the corresponding eigenvector. [All such faces are maximal and isometric to
${\cal M}^{(N-1)}$.] The
point  of tangency of that face to the inscribed (Hilbert--Schmidt) ball will
be a density matrix with one eigenvalue (corresponding to the same eigenvector
as above) zero and all other eigenvalues equal.
[Like the set of pure states, both the set of maximal faces and the set $Y$ of
points of tangency have the structure of a complex projective space.]  Of
course, we can also derive the conclusion of Proposition 1 from (\ref{muMN})
and the condition
${\bf 3}$ of Lemma 1, but such an argument would not be self-contained and,
moreover, would obscure the matter.

\medskip

\begin{figure} [htbp]
       \begin{center} \
     \includegraphics[width=10.5cm,angle=0]{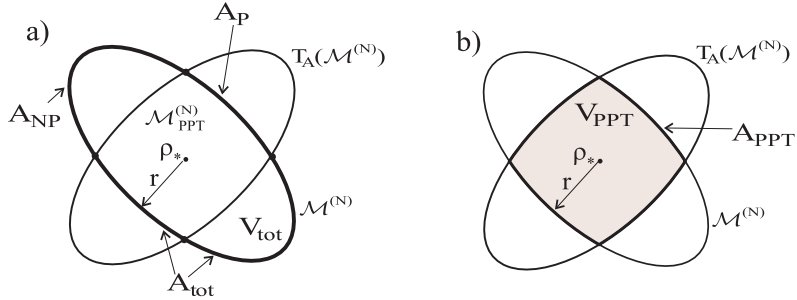}
\caption{The set ${\cal M}^{(N)}_{PPT}$ of PPT states
arises as a common part of
the set of mixed states ${\cal M}^{(N)}$
and its image $T_A({\cal M}^{(N)})$
with respect to partial transpose. Its volume and area are denoted by
$V_{\rm PPT}$ and $A_{\rm PPT}$.}
\label{fig1}
\end{center}
     \end{figure}

Let us now take the dimension $N$ as a composite number, say $N=KM$,
and consider operators acting on a composite Hilbert space
${\cal H}_N={\cal H}_K\otimes {\cal H}_M$. Such a decomposition
of the Hilbert space allows one to define {\it separable} states, as those
represented as a convex sum of product states \cite{We89},
\begin{equation}
\rho_{sep}\: = \: \sum_{j=1}^L \, q_j \, \rho^A_j \otimes \rho^B_j \ ,
     \label{mixedsep}
\end{equation}
where operators $\rho^A$ and $\rho^B$ act on Hilbert spaces
${\cal H}_K$ and ${\cal H}_M$, respectively,
the weights are positive, $q_j>0$,  and  sum to unity, $\sum_{j=1}^L q_j=1$.
A state which is not separable  is called {\it entangled}.
Any separable state has a positive partial transpose,
$T_A(\rho)=(T  \otimes {\mathbbm 1})\rho \ge 0$,
and this criterion is sufficient if $N=2 \times 2=4$ or $N=2 \times 3=6$
\cite{Pe96,HHH96a}. Thus the sets of separable states and  PPT states
(states with positive partial transpose) coincide in these cases
and are equal to the intersection of ${\cal M}^{(N)}$
and its image $T_A({\cal M}^{(N)})$ -- see Fig. \ref{fig1}.
In any dimension $N = KM$ the sets ${\cal M}^{(N)}$
and $T_A({\cal M}^{(N)})$ have the same shape, volume and surface
area, because the partial transpose $T_A$ acts as a reflection with
respect to the affine subspace of PPT--invariant states (which includes
the maximally mixed state $\rho_*$).

\smallskip
Let us denote the volume and the area
of the set of mixed states by
$V_{\rm tot}={\rm Vol}({\cal M}^{(N)})$ and
$A_{\rm tot}={\rm Vol}(\partial {\cal M}^{(N)})$.
Analogously
$V_{\rm PPT}$ and $A_{\rm PPT}$ represent
the volume and the area of the set of PPT states.
The key result of this work will follow from

\smallskip

\noindent
{\bf Theorem 1.} {\sl The set ${\cal M}^{(N)}_{PPT}$ of PPT mixed
quantum states is
a body of constant height.}
\smallskip

\noindent {\it Proof:} The set ${\cal M}^{(N)}$ of mixed states is a body of
constant height, and so is its image, $T_A({\cal M}^{(N)})$. Both
sets contain the same inscribed sphere of radius
$r=1 / \sqrt{(N-1)N}$.
Since the set of PPT states is an intersection
of two bodies of constant height,
${\cal M}^{(N)}_{PPT}={\cal M}^{(N)} \cap T_A({\cal M}^{(N)})$,
the intersection lemma (Lemma 2) implies that this set has constant height.
\hfill $\square$
\smallskip

The total area of the set ${\cal M}^{(N)}$ of mixed states can be
divided into two parts: one part $A_P$ containing PPT states, and
another part $A_{NP}$ containing not PPT states -- see Fig. \ref{fig1}a.
By definition
$A_{\rm tot}=A_{NP}+A_P$.
On the other hand, the surface of the set of
PPT states is a union of two congruent parts:  the part $A_P$
described above, and its isometric image under $T_A$
-- see Fig. \ref{fig1}b.
Accordingly, it is possible to infer that the area $A_{\rm PPT}$
of the set ${\cal M}^{(N)}_{PPT}$ equals
$2 A_P$ once we establish that the area of the intersection of
the two parts (the ``corners" of the set PPT on Fig. \ref{fig1}) is
zero. 
This is indeed the case and
is the content of the following
\smallskip

\noindent
{\bf Lemma 3.}  {\sl  The $D-1$-dimensional area of the intersection
of the boundary $\partial {\cal  M}^{(N)}$ with its image under $T_A$
is zero}.

\smallskip
\noindent
Before we prove this, we need to recall some basic facts about
the boundary $\partial {\cal  M}^{(N)}$.
First of all any density matrix can be written as

\begin{equation} \rho = UE U^{\dagger} \ , \label{orbits} \end{equation}

\noindent where $U$ is a unitary matrix and $E$ is diagonal. 
As indicated in the comments
following Proposition 1, $\partial {\cal M}^{(N)}$ consists exactly of
those density matrices that have a zero eigenvalue. Further,
after disregarding the density
matrices with multiple eigenvalues (a subset of dimension
$D-2$, and hence of surface measure zero), the remaining
``generic" part of $\partial {\cal  M}^{(N)}$ is a smooth manifold
which---because of eq.
(\ref{orbits})---is naturally diffeomorphic to the
Cartesian product of the open simplex
$\Lambda := \{(0,\lambda_2,\ldots, \lambda_N) : 0<\lambda_2<\ldots<
\lambda_N, \sum_{j=2}^N \lambda_j = 1\}$ and the
flag manifold $Fl_N^{\mathbb C} = U(N)/U(1)^N$. The first component is
just the ordered sequence of eigenvalues of the matrix, while the second
component is the set of of unitary matrices with those matrices that
commute with $E$ ``divided out''; the second component is in fact
given by the corresponding sequence of eigenspaces.
(A more detailed analysis of the stratification of  ${\cal M}^{(N)}$
can be found in \cite{GMK05}.) Moreover, as is well-known (cf., e.g.,
\cite{ZS03}), this diffeomorphism transforms the surface measure
to a product measure of the type
$f(\lambda) \, d\lambda \otimes du$, where $d\lambda$ is the Lebesgue
measure on $\Lambda$, $f(\lambda)$ -- a positive continuous function
on $\Lambda$, and $du$ is the invariant measure on $Fl_N^{\mathbb C}$
induced by the action of $U(N)$.

The heuristic idea behind the proof that follows is that the area 
of intersection is zero if the normals of the two hypersurfaces are distinct 
in ``nearly all'' places where they intersect. However, some care is needed 
in order to ensure that this argument works.

\noindent {\it Proof of Lemma 3:} 
We will first prove a weaker statement, namely that
the intersection
$\partial {\cal  M}^{(N)} \cap T_A(\partial {\cal  M}^{(N)})$
has empty interior (relative to $\partial {\cal  M}^{(N)}$).

We shall argue by contradiction. Assume that the
intersection {\em does have} nonempty interior. Then
the subset $C_{gen}$ of the intersection consisting
of density matrices $\rho$ verifying
\smallskip

\noindent
   $\bullet$ $\rho$ and $T_A(\rho)$ belong to the
generic part of
$\partial {\cal  M}^{(N)}$
\smallskip

    \noindent
$\bullet$ the pure states corresponding to the
eigenvectors of
$\rho$ and $T_A(\rho)$ are {\em not} separable

\smallskip

\noindent also has a nonempty interior  since each of
the above two conditions is satisfied
outside a (closed) set of dimension $<D-1$.
Let  $\rho_0$ be an interior point of $C_{gen}$.
We now recall that the first
eigenvector (of eigenvalue zero) of a generic boundary density matrix
$\rho$ corresponds to the pure state which is normal to
$\partial {\cal M}^{(N)}$ at $\rho$. Thus, by our assumption, the normal to
$\partial {\cal M}^{(N)}$ at $T_A(\rho_0)$ corresponds to an entangled
state and, consequently, its image under $T_A$ -- which is the normal to
$T_A(\partial {\cal M}^{(N)})$ at $\rho_0$ -- corresponds to a trace one
matrix which is not positive-definite.
(This is because for all entangled {\em pure } states
their negativity is positive \cite{VW02,ZB02},
so the PPT criterion is necessary and sufficient for separability
of pure states).
    In particular, the (unique) normals to $T_A(\partial
{\cal M}^{(N)})$ and $\partial {\cal M}^{(N)}$ at $\rho_0$ are
{\em distinct}. This implies that, in sufficiently small
neighborhoods of $\rho_0$, the intersection of $T_A(\partial
{\cal M}^{(N)})$ and $\partial {\cal M}^{(N)}$ is a
$D-2$-dimensional object and so $\rho_0$ cannot be an interior point of the
intersection and {\em a fortiori} of $C_{gen}$, a contradiction.

Since $\partial {\cal  M}^{(N)} \cap T_A(\partial {\cal  M}^{(N)})$ is
closed,
the above argument does shows that  it is a rather ``thin" set
(``nowhere dense" in topological parlance). However,
this is not sufficient for our purposes because intersections of convex
surfaces may be rather complicated, in particular they may be of positive
area in spite of having empty (relative) interior. To correct this
shortcoming,
we need to choose $\rho_0 \in C_{gen}$ to be a Lebesgue point of
(the indicator  function of) $C_{gen}$ rather than an interior point of
$C_{gen}$.
Then, for every sufficiently regular sequence $(N_k)$ of neighborhoods of
$\rho_0$  (relative to $\partial {\cal M}^{(N)}$) which
shrinks to $\rho_0$, the ratios of the areas of $N_k \cap
C_{gen}$ and of $N_k$ tend to 1 (this is  just a reformulation
of  the definition of a Lebesgue point).  By the Lebesgue differentiation
theorem, almost all points of a measurable set have this property (see,
e.g.,
\cite{folland}, Theorem 3.21). Accordingly, if we assume that
$\partial {\cal  M}^{(N)} \cap T_A(\partial {\cal  M}^{(N)})$, hence
$C_{gen}$,
has strictly positive area,  then such a
  choice of $\rho_0 \in C_{gen}$  is possible and we are led to a
  contradiction
  in the same way as earlier.\hfill
$\square$

\medskip

The ratio between the probabilities
of finding a PPT state in the interior of ${\cal
M}^{(N)}$ and at its boundary
is defined by  \cite{Sl05a},

\begin{equation}
\Omega \equiv \frac{p_V} {p_A} :=
    \frac{V_{\rm
PPT}/ V_{\rm tot}} {A_P / A_{\rm tot}}  =
\frac{V_{\rm PPT} \,
A_{\rm tot}} {V_{\rm tot} \, A_{P}}  \ .
\label{Omega1}
\end{equation}
The above quantity allows us to formulate
the main result of the work
\smallskip

\noindent
{\bf Theorem 2.}  {\sl  For any bipartite $K \times M$ system
the ratio} (\ref{Omega1}) {\sl  is equal to $2$.}

\smallskip

\noindent {\it Proof:} By Theorem 1 the set of PPT states has constant height,
hence
${\gamma}({\cal M}^{(KM)}_{PPT})= r A_{\rm PPT}/ V_{\rm PPT}=D$.
For any fixed system of size $N=KM$,
the sets ${\cal M}^{(N)}$ and ${\cal M}_{\rm PPT}^{(N)}$
have the same dimensionality $D$ and the same radius $r$ of the inscribed ball,
so that
\begin{equation}
\frac{A_{\rm tot}} {V_{\rm tot}} \ = \ \frac{A_{\rm PPT}}{V_{\rm PPT}} \ .
\label{AVAV}
     \end{equation}
Lemma 3 implies $A_P=A_{\rm PPT}/2$.
Substituting this relation to (\ref{Omega1})
and making use of (\ref{AVAV}) we find
\begin{equation}
\Omega  =
\frac{V_{\rm PPT}\, A_{\rm tot}} {V_{\rm tot}\, A_{\rm PPT}/2} \: =
\: 2 \ .
\label{Omega2}
    \end{equation}
which concludes the proof. \hfill
$\square$
\smallskip

The relation $\Omega=2$ is thus established
for any bipartite system. In the simplest cases of $2 \times 2$
and $2 \times 3$ systems any PPT state is separable  \cite{HHH96a}.
Hence, for such systems, eq. (\ref{Omega2}) describes the ratio
between the probabilities of finding a separable state inside
the set of mixed states and at its boundary.

The value
$\Omega=2$ is consistent with the numerical data obtained in
\cite{Sl05a,Sl05b} for the Hilbert--Schmidt measure.
Since our reasoning hinges directly
on the Euclidean geometry, it does not allow to predict any
values of analogous ratios computed with respect
to Bures measure \cite{SZ03}, nor other measures.
On the other hand, the
result obtained is valid also for the
set of real density matrices,
for which the complex flag manifold $Fl_N^{\mathbb C}$
has to be replaced by the real flag manifold,
$Fl_N^{\mathbb R} = O(N)/O(1)^N$.
The set of real density matrices
is often used as an attractive toy model,
since its dimensionality $D_R=N(N+1)/2-1$
is smaller than $D=N^2-1$ of the full
set of complex states.

\medskip
The geometry of
the $15$-dimensional set of separable states of
two qubits is not easy to describe. In this work we have
established a concrete
rigorous result in this direction, proving that this set is
``pyramid-decomposable" and hence is a body of constant height.
This is also true for the set of positive-partial-transpose
states for an arbitrary bipartite system.
We hope that these properties will be useful
for further investigation of
the geometry of the set of separable states.
Although in this work we have concentrated
on the bipartite case only,
one could try to obtain similar results for a
general class of multipartite systems,
for which some estimates of
the volume of the set of
separable states or PPT states are
known \cite{Sza04,AS05,GB03,GB04}.

\medskip

We would like to thank
Paul Slater for numerous
discussions during the past years and
letting us know
about his results. Without his input this work would
never
have been started. I.B. and K.{\.Z}. acknowledge hospitality
of the Perimeter Institute for Theoretical Physics in Waterloo,
where this work was initiated.
K.{\.Z}. is grateful for a partial support by
Polish Ministry of Science and Information Technology
under the grant PBZ-MIN-008/P03/2003.
S.J.S. has been partially supported by
the National Science Foundation (U.S.A.).


\end{document}